\documentclass[floatfix,aps,nofootinbib]{revtex4}
\usepackage{graphicx}

\begin{document}

\title{The energy scale of inflation: is the hunt for the primordial
       B-mode a waste of time?}
\author{William H.\ Kinney} \email{kinney@physics.columbia.edu}
\affiliation{Institute for Strings, Cosmology and Astroparticle Physics,
	Columbia University, 550 W. 120th St., New York, NY 10027}
\date{1 July 2003}

\begin{abstract}
Recent theoretical results indicate that the detection of primordial
gravity waves from inflation may be a hopeless task. First, foregrounds
from lensing put a strict lower limit on the detectability of the B-mode
polarization signal in the Cosmic Microwave Background, the ``smoking gun''
for tensor (gravity wave) fluctuations. Meanwhile, widely accepted
theoretical arguments indicate that the amplitude of gravity waves produced
in inflation will be below this limit. I argue that failure is not
inevitable, and that the effort to detect the primordial signal in
the B-mode, whether it succeeds or fails, will yield crucial information
about the nature of inflation.
\end{abstract}

\maketitle

\section{Introduction}

\begin{figure}
\centerline{\includegraphics[width=4.5in]{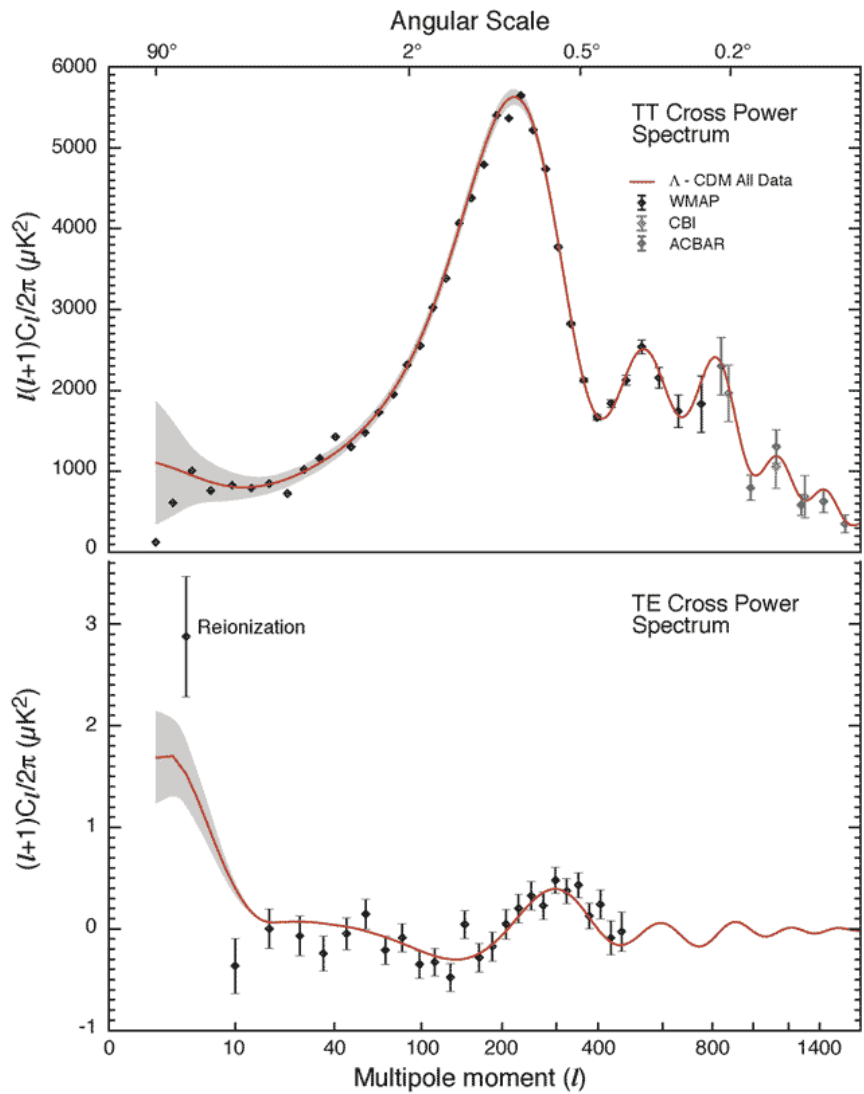}}
\caption{\label{WMAPspectrum}Angular power spectra observed by WMAP. The
top panel shows the temperature anisotropy, and the bottom panel shows
the temperature-polarization cross correlation spectrum. (Figure courtesy
of the WMAP science team.) Especially important for the confirmation of
acausal physics typical of inflation is the
measured T/E anticorrelation around $\ell = 100$.}
\end{figure}

Now that the era of precision cosmology is a reality, inflation is enjoying
a period of stunning success . Inflation has not only
been successful in its broad-brush prediction of a flat universe, but also
in its more detailed prediction of a Gaussian, adiabatic,
nearly (but not exactly) scale-invariant spectrum of primordial perturbations.
Observations of the Cosmic Microwave Background (CMB) have been especially
instrumental in confirming the predictions of inflation. In particular,
the WMAP satellite has painted a picture of the universe that is
precisely consistent with that expected from inflation
\cite{peirisetal,barger03,kkmr,leach03}. 
Figure \ref{WMAPspectrum}
shows the angular power spectra of CMB fluctuations observed by WMAP
\cite{WMAPbasic}.
Particularly exciting is that WMAP did not just measure the anisotropy in
the temperature of the CMB, but also its {\em polarization}. The WMAP
polarization measurement provides us with what is probably the least
ambiguous signal for inflationary physics: a measured anticorrelation
between the polarization and the temperature fluctuations on an angular
scale around $\ell = 100$, evident in the bottom panel of Fig. 1. This
feature is important because it is an essentially foreground-free
measurement of correlations in primordial fluctuations on scales larger
than the horizon size, a ``smoking gun'' for the acausal physics
characteristic of inflation \cite{zaldarriaga97}.
(Large-angle correlations in the temperature do not provide a clean
signal of such acausal physics because of the possibility of foregrounds
such as the integrated Sachs-Wolfe effect.)

Future CMB observations will improve considerably on existing data sets,
especially in the measurement of the polarization signal. CMB polarization
spectra divide into two types depending on their parity under reflections
of the celestial sphere: even-parity, or ``E-modes'', and odd-parity,
or ``B-modes''. The E-mode was first detected by DASI \cite{DASIemode}, and
the temperature / E-mode cross correlation was measured by
WMAP. The B-mode signal is much smaller, and has yet to be detected. The
B-mode is of particular interest for learning about inflation, because
it is the only CMB signal that receives no contribution from primordial
density fluctuations. Instead, the B-mode is generated entirely by
primordial gravitational wave fluctuations. Since both scalar (density)
and tensor (gravitational wave) fluctuations are generated during inflation,
detection of tensor modes would greatly increase our ability to place
constraints on the inflationary model space. In particular, the amplitude
of the tensor fluctuations (unlike scalars) depends only on the value of
the Hubble constant during inflation, or equivalently the potential of
the scalar field driving inflation (the {\em inflaton}):
\begin{equation}
P_{\rm T} = \left({H \over 2 \pi m_{\rm Pl}}\right)^2 =
{2 V\left(\phi\right)\over 3 \pi m_{\rm Pl}^4}.
\end{equation}
Therefore, if we measure the amplitude of primordial tensor fluctuations,
we can determine the energy scale of inflation. This makes the B-mode
polarization an important observational target for cosmology. The big
question is then, how large a B-mode do we expect to be generated by a
typical inflation model?

\section{Effective field theory and the energy scale of inflation}

The prevailing prejudice in model building is that the inflaton is
a fundamental field: a scalar
degree of freedom in some low-energy limit of an underlying, fundamental
theory such as supergravity or string theory. Therefore we expect the
techniques of effective field theory, for which heavy degrees of freedom
are integrated out, to be a correct description of the physics of
inflation. In other words, expect the effective potential for the inflaton
to be an expansion in nonrenormalizable operators suppressed by some
higher energy scale, which we take to be the Planck mass:
\begin{equation}
\label{eqeffectivepotential}
V\left(\phi\right) = V_0 + {1 \over 2} m^2 \phi^2 + \phi^4
\sum_{p = 0}^{\infty}{\lambda_p \left(\phi \over m_{\rm Pl}\right)^p}.
\end{equation}
The effective Lagrangian for the inflaton will also in general contain
corrections (also suppressed by the Planck mass) to kinetic terms, but
we need not consider these for the purpose of the current argument. The
important feature of this construction is that the effective theory is
only self-consistent for $\phi << m_{\rm Pl}$, otherwise the series
expansion for the effective potential (\ref{eqeffectivepotential}) will
not in general be convergent. The question we wish to answer here is:
how well does such a construction work in the context of inflation?

To illustrate, we consider the simple case of a monomial ``small field''
potential with height characterized by a scale $\Lambda$ and width
characterized by a scale $\mu$,
\begin{equation}
\label{eqgenericsmallfieldV}
V\left(\phi\right) = \Lambda^4 \left[ 1 -
\left({\phi \over \mu}\right)^p\right].
\end{equation}
For potentials of this form, inflation takes place when the field value
is small, $\phi \ll \mu$, and the slope of the potential is also small,
$V' \ll V$ so that the field is slowly rolling and the contribution
of the potential to the energy density of the field dominates over the
contribution of the kinetic energy. Eq. (\ref{eqgenericsmallfieldV})
represents the form of the potential near its maximum $\phi = 0$.
Inflation ends and reheating takes place when the field value is comparable
to the width parameter $\phi \sim \mu$. So the field travels a distance
$\Delta \phi \sim \mu$ during inflation. Therefore our assumption of a valid
effective field theory expansion (\ref{eqeffectivepotential}) is only
consistent if $\Delta \phi \sim \mu \ll m_{\rm Pl}$ \cite{lyth96}. We first
take the case $p = 2$. The spectral index of scalar fluctuations can be
calculated to be
\begin{equation}
n = 1 - {1 \over 4 \pi} \left({m_{\rm Pl} \over \mu}\right)^2.
\end{equation}
We see that the scale-invariant limit is reached for
$\Delta \phi \sim \mu >> m_{\rm Pl}$. For $n > 0.95$, we must have
$\mu > 1.3 m_{\rm Pl}$, and we expect effective field theory to break
down at some point during the inflationary evolution. Such potentials
predict a tensor/scalar ratio $r \equiv (C^T_2 / C^S_2) \simeq 0.01$
for $n \simeq 0.95$. This
is not the most extreme example. For a potential of the form
$V\left(\phi\right) = \lambda \phi^4$, which predicts a large tensor/scalar
ratio $r \sim 0.1$, the field travels $\Delta \phi \sim 4  m_{\rm Pl}$
during inflation.
However, consider a potential of the form (\ref{eqgenericsmallfieldV}) with
$p = 4$. The spectral index of scalar fluctuations is given by $n \simeq 0.95$
{\em regardless} of the values of the fundamental scales $\Lambda$ and $\mu$
in the potential \cite{kinney95}. So, unlike the case of the quadratic
potential, it is perfectly consistent with
observation to have a potential with $\Delta \phi \sim \mu \ll m_{\rm Pl}$.
However, one does this at the price of having a vanishingly small
tensor/scalar ratio, $r \ll 10^{-3}$.

Lyth \cite{lyth96} showed this to be true in general. Subject to
the condition of a nearly scale-invariant spectrum, the height of the
potential can be written in terms of the tensor/scalar ratio $r$ as
approximately
\begin{equation}
\Lambda \sim \left({r \over 0.7}\right)^{1/4} \times 1.8
\times 10^{16}~{\rm GeV}.
\end{equation}
This is the familiar result that a significant contribution of tensor
fluctuations to the CMB requires that inflation take place at a high
energy scale. More important for the discussion here is that the
{\em width} of the potential can also be related to the tensor/scalar
ratio
\begin{equation}
{\Delta \phi \over m_{\rm Pl}} \sim 0.5 \left({r \over 0.07}\right)^{1/2}.
\end{equation}
This means that for a tensor/scalar ratio of order $0.1$, the field
must travel a distance $\Delta \phi > m_{\rm Pl}$ during inflation, so
that the effective field theory expansion (\ref{eqeffectivepotential})
must break down at some point during inflation. An effective potential
that works near the beginning of inflation will be divergent near the
end of inflation, and the entire description in terms of an effective
Lagrangian is invalid. This fact has been used as an argument, based on
self-consistency, for expecting a very small tensor/scalar ratio in a
realistic inflationary universe.

This is a discouraging conclusion, because it means that the signature of
primordial gravitational waves, in particular the B-mode component
of the CMB, will be unobservably small. Knox and Song \cite{knox02} have
shown that foregrounds from gravitational lensing by cosmological structure
place a fundamental lower limit on how well the B-mode can be measured,
corresponding to a lower limit on the tensor/scalar ratio of about
\begin{equation}
r > 6 \times 10^{-4},
\end{equation}
which corresponds to an energy scale for inflation of roughly
\begin{equation}
\Lambda > 3.2 \times 10^{15}\ {\rm GeV}.
\end{equation}
If the energy scale of inflation is below this limit, the B-mode (and hence
the primordial gravitational wave background) will be unobservable in the
CMB. (A recent paper \cite{hirata03} indicates that it may be possible to
use a more optimal estimator to improve the Knox and Song limit by as much
as a factor of two. This does not substantially alter the discussion here.)

\section{The flow approach to inflationary evolution}

\begin{figure}
\centerline{\includegraphics[width=4.5in]{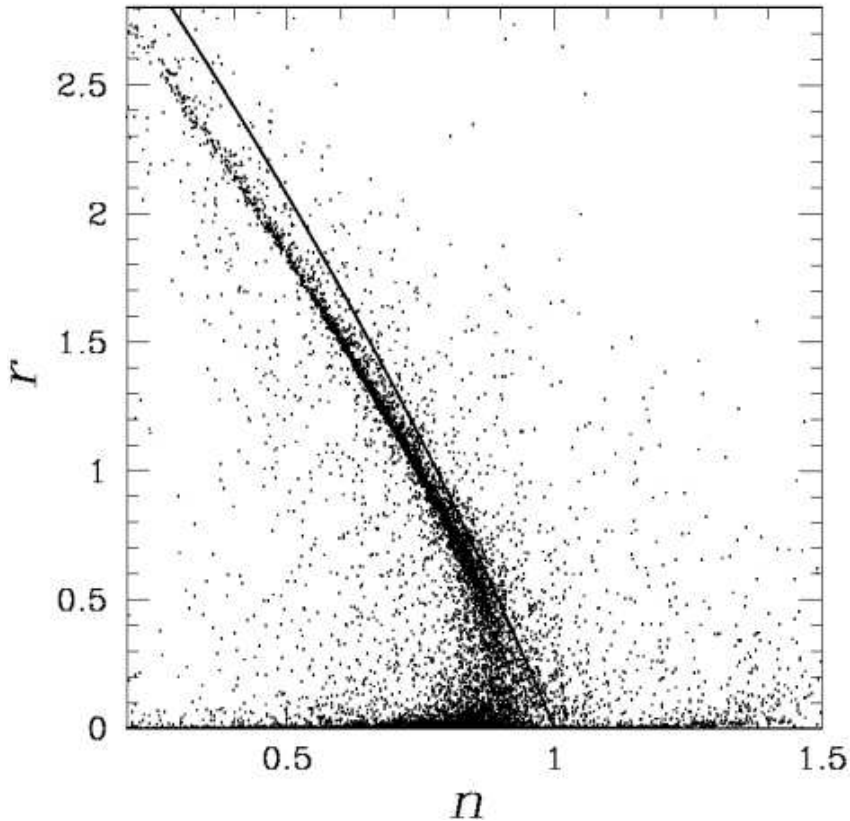}}
\caption{\label{nrMC}Models generated by Monte Carlo plotted in the $(n,r)$ plane.
The solid line is the power-law inflation fixed point $n = 1 - 2 r / (10 - r)$.}
\end{figure}

\begin{figure}
\centerline{\includegraphics[width=4.5in]{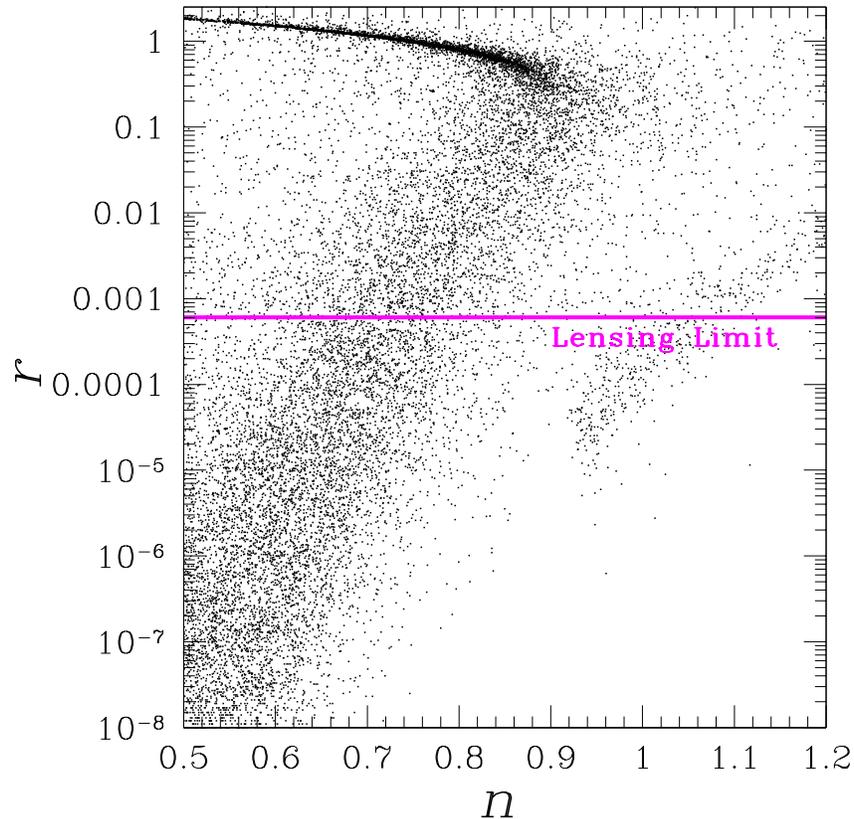}}
\caption{\label{nlogrMC} n vs log r. The horizontal line is the lensing limit
of Knox and Song. There is a substantial population of models with a large
enough tensor/scalar ratio to be detectable above the lensing foreground.}
\end{figure}

Arguments based on effective field theory, however, are not the only way
to look at the dynamics of inflation. The inflaton, for example, may not
be a fundamental field -- all that is necessary for the predictions of
single ``field'' inflation to be valid is that the evolution of the
spacetime be governed by a single order parameter. It is desirable to
look at the predictions of inflation without resorting to assumptions
about effective field theory. One way to do this is to reformulate the
dynamical equations for inflation as flow in the space of dimensionless
slow roll parameters \cite{flowpapers}. A major advantage of this approach is
that it removes the field from the dynamical picture altogether, so
that we can study the generic behavior of slow roll inflation without
making assumptions about the nature of the underlying particle physics. To
construct the hierarchy of flow equations, we start with the slow roll
parameters
\begin{equation}
\epsilon \equiv {m_{\rm Pl}^2 \over 4 \pi} \left({H'\left(\phi\right)\over
H\left(\phi\right)}\right)^2,
\end{equation}
and
\begin{equation}
\eta \equiv {m_{\rm Pl}^2 \over 4 \pi} \left({H''\left(\phi\right)\over
H\left(\phi\right)}\right).
\end{equation}
To lowest order, the observables $r$ and $n$ are given to lowest order
in slow roll by\footnote{The normalization of $r$ used here differs from
that used by the WMAP team, who define $r \simeq 16 \epsilon$ \cite{peirisetal}.}
\begin{eqnarray}
\label{eqobservables}
r &\simeq& 10 \epsilon,\cr
n - 1 &=& 4 \epsilon - 2 \eta \equiv \sigma.
\end{eqnarray}
Here we have introduced $\sigma \equiv 4 \epsilon - 2 \eta$ for notational
convenience. These can be seen as the first members of an infinite hierarchy
of slow roll parameters \cite{liddle94}
\begin{equation}
{}^\ell\lambda_{\rm H} \equiv \left({m_{\rm Pl}^2 \over 4 \pi}\right)^\ell
 {\left(H'\right)^{\ell-1} \over H^\ell} {d^{(\ell+1)} H \over d\phi^{(\ell +
 1)}}.\label{eqdefoflambda}
\end{equation}
Each of these parameters can be written as a function of the number of
e-folds $N$ before the end of inflation, where
\begin{equation}
{d \over d N} = { m_{\rm Pl} \over 2 \sqrt{\pi}} \sqrt{\epsilon} {d \over
 d\phi},
\end{equation}
and we have an infinite series of differential flow equations
describing the inflationary evolution
\begin{eqnarray}
{d \epsilon \over d N} &=& \epsilon \left(\sigma + 2 \epsilon\right),\cr
{d \sigma \over d N} &=& - 5 \epsilon \sigma - 12 \epsilon^2 + 2
 \left({}^2\lambda_{\rm H}\right),\cr
{d \left({}^\ell\lambda_{\rm H}\right) \over d N} &=& \left[{1 \over 2}
 \left(\ell - 1\right) \sigma + \left(\ell - 2\right) \epsilon\right]
 \left({}^\ell\lambda_{\rm H}\right) + {}^{\ell+1}\lambda_{\rm
 H}.\label{eqfullflowequations}
\end{eqnarray}
The field has disappeared from the equations of motion altogether, although
the underlying assumption of dynamics controlled by a single order parameter
is of course still present. The flow equations allow us to investigate the
model space for inflation using Monte Carlo techniques. Since the
dynamics are governed by a set of first-order differential equations, the
cosmological evolution is entirely specified by choosing values for the
slow roll parameters $\epsilon, \sigma, {}^2\lambda_{\rm H},\ldots$. Choosing
such a point in the parameter space completely specifies the inflationary
model, including the scalar field potential, which can be reconstructed
for each choice, so-called ``Monte Carlo reconstruction'' \cite{montecarlorecon}.
The observable predictions for a given model can be evaluated as follows:
\begin{itemize}
\item{Pick a point in the parameter space $\epsilon, \sigma,
{}^2\lambda_{\rm H},\ldots$}
\item{Evolve forward in time until inflation ends, or reaches a late-time
attractor.}
\item{From the end of inflation, evolve backward in time about 60 e-folds,
and calculate the values of the slow roll parameters at that point.}
\item{Calculate the observables $r$, $n$, $dn/d\ln{k}$.}
\end{itemize}
This procedure can be performed numerically for a large number of randomly
chosen initial conditions. Instead of specifying a potential and calculating
the predictions for that model (``hand crafting'' inflation), we can
process models on an industrial scale, investigating the behavior of millions
of possibilities for the inflationary dynamics. The result is very interesting:
models do not uniformly cover the observable parameter space, but instead
cluster around attractor regions. How well do the attractors correspond to
the expectations from effective field theory? Not especially closely. Figure
\ref{nrMC} shows the models plotted in the $r,n$ plane, showing that there
is a significant concentration of models with nonzero tensor/scalar ratio
$r$.
Figure \ref{nlogrMC} shows the same points plotted logarithmically in
tensor/scalar ratio, with the lower bound from Knox and Song marked on the
plot. There is a substantial population of dynamically valid inflation
models with a tensor scalar ratio that is {\em in principle} observable,
and there is no compelling reason to rule such models out. Observation
of a significant tensor/scalar ratio would be a strong indication of
some kind of exotic physics driving inflation, which is not adequately captured
by a description in terms of a low-energy effective field theory with
operators suppressed by powers of the Planck scale. Such a result would be
of great interest from the standpoint of inflationary model building, but
would not in itself weaken inflation as a viable theory.

\section{The current observational situation and future prospects}

\begin{figure}
\centerline{\includegraphics[width=4.2in]{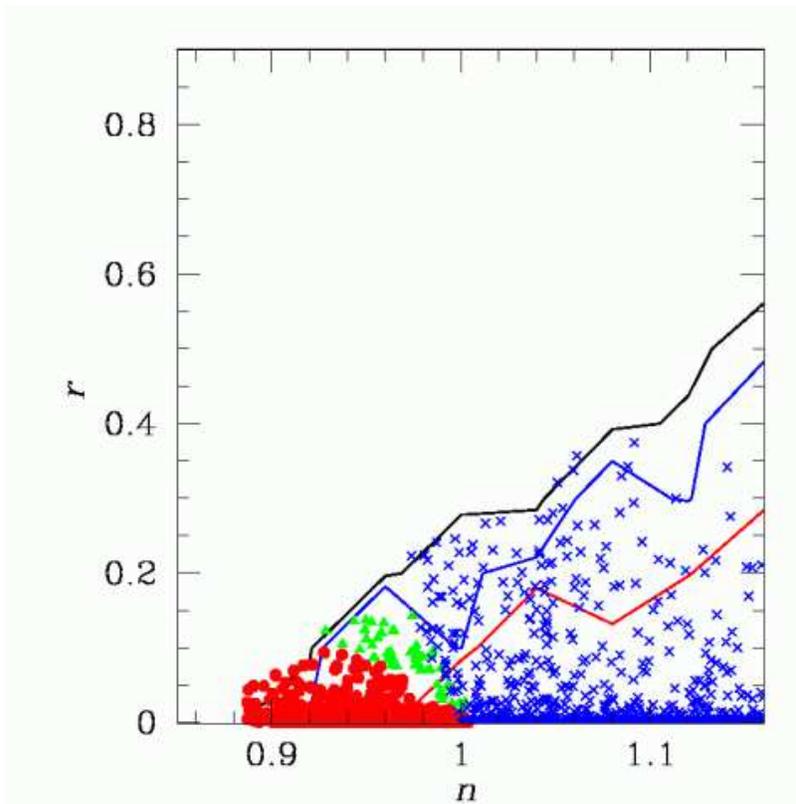}}
\caption{\label{nrWMAP} Errors from current CMB data on the $(r,n)$ plane. The
contours represent $1\sigma$, $2\sigma$, and $3\sigma$ errors. The points are
models generated by Monte Carlo, categorized into small-field (red, circles),
large-field (green, triangles), and hybrid (blue, crosses). No class of models
is yet ruled out by the data.}
\end{figure}

\begin{figure}
\centerline{\includegraphics[width=4.2in]{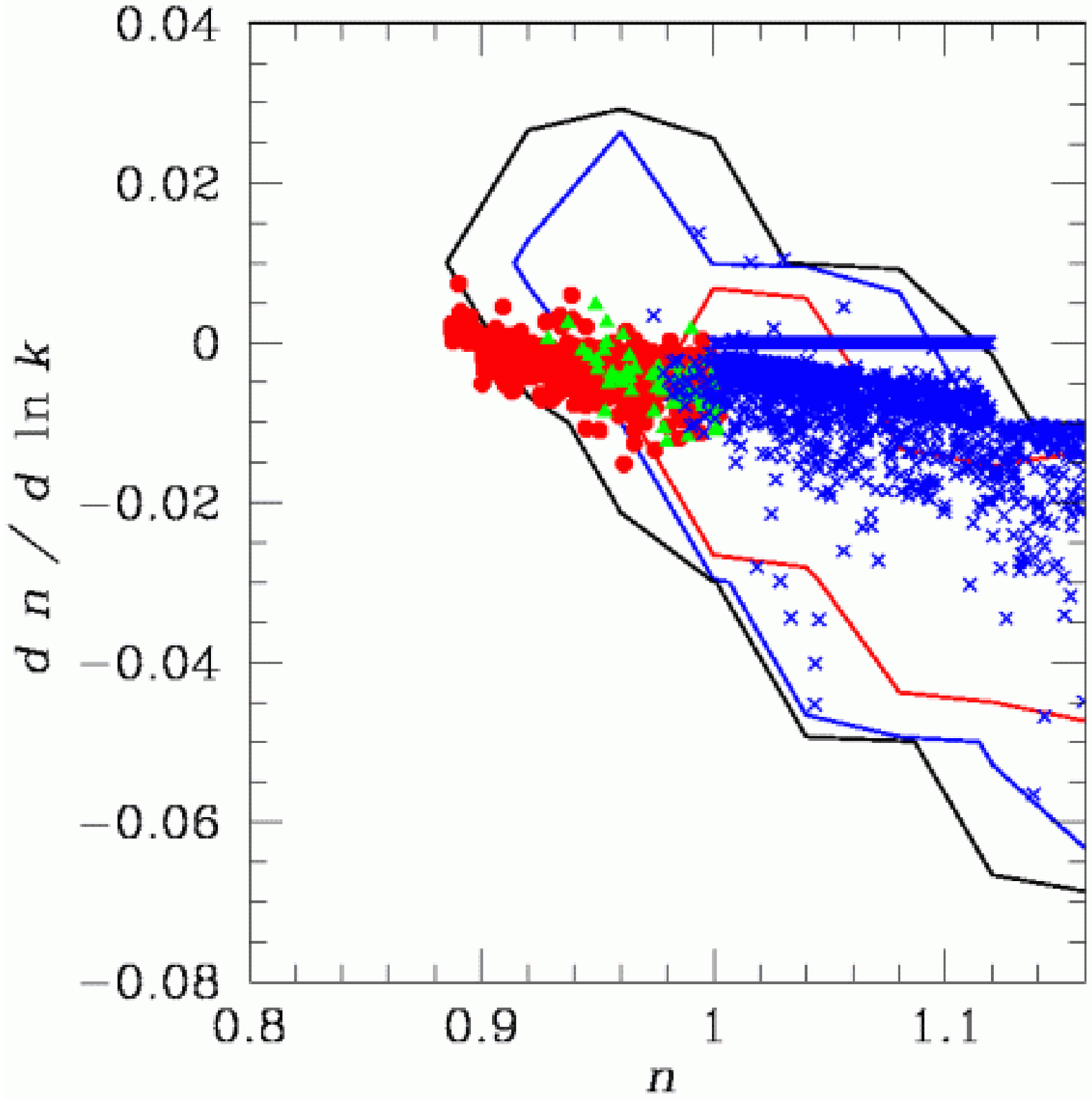}}
\caption{\label{ndnWMAP} Errors from current CMB data on the $(n, dn/d\log{k})$
plane. Contours represent $1\sigma$, $2\sigma$, and $3\sigma$ errors, with
the models plotted as in Fig. \ref{nrWMAP}. Note that the region with negative
running is well populated by models, especially of the hybrid class, indicating
that the WMAP best fit is easily accomodated by simple inflationary models.}
\end{figure}

What are the prospects in the near future for detection of a tensor contribution
to the CMB? Certainly, it is not likely that we will be able to approach the
lensing bound on the B-mode any time soon. Current limits on the size of
the tensor contribution to the CMB are still quite crude, although the
recent WMAP result represents a substantial step forward.
Figure \ref{nrWMAP} \cite{kkmr} shows the constraint on the $n,r$ plane from the WMAP data
set in conjunction with seven additional CMB data sets (BOOMERanG-98~\cite{ruhl},
MAXIMA-1 \cite{lee}, DASI~\cite{halverson}, CBI~\cite{cbi}, ACBAR~\cite{acbar},
VSAE~\cite{grainge}, and Archeops~\cite{benoit}). Figure \ref{ndnWMAP} \cite{kkmr} shows
the constraint in the $n,dn/d\log{k}$ plane, showing a preference for a negative
running of the spectral index with scale, consistent with the results of the
data analysis done by the WMAP team\cite{peirisetal}. The points plotted are
models generated by Monte Carlo evaluation of the flow equations consistent with
the data to $3\sigma$, and are coded by their ``zoology'': small field, large field,
and hybrid.\footnote{See Ref. \cite{kkmr} for a detailed discussion of this categorization.}
We see that while no class of models is ruled out by the data, the
hybrid class shows the greatest consistency with the best-fit region. There is
in fact a population of hybrid models with a strong negative
running of the spectral index, showing that the WMAP best fit is quite consistent
with what one might expect from inflation. Deviations from the limiting case
of a scale-invariant power-law spectrum are exactly what one ought to expect
in a realistic inflationary cosmology: astrophysical observations probe the
last sixty e-folds, i.e. the {\em end} of inflation, when the limiting behaviors
of slow roll are beginning to break down. In terms of the current data, inflation
is in excellent shape as a model for production of the primordial density
fluctuations.

\begin{figure}
\centerline{\includegraphics[width=4.2in]{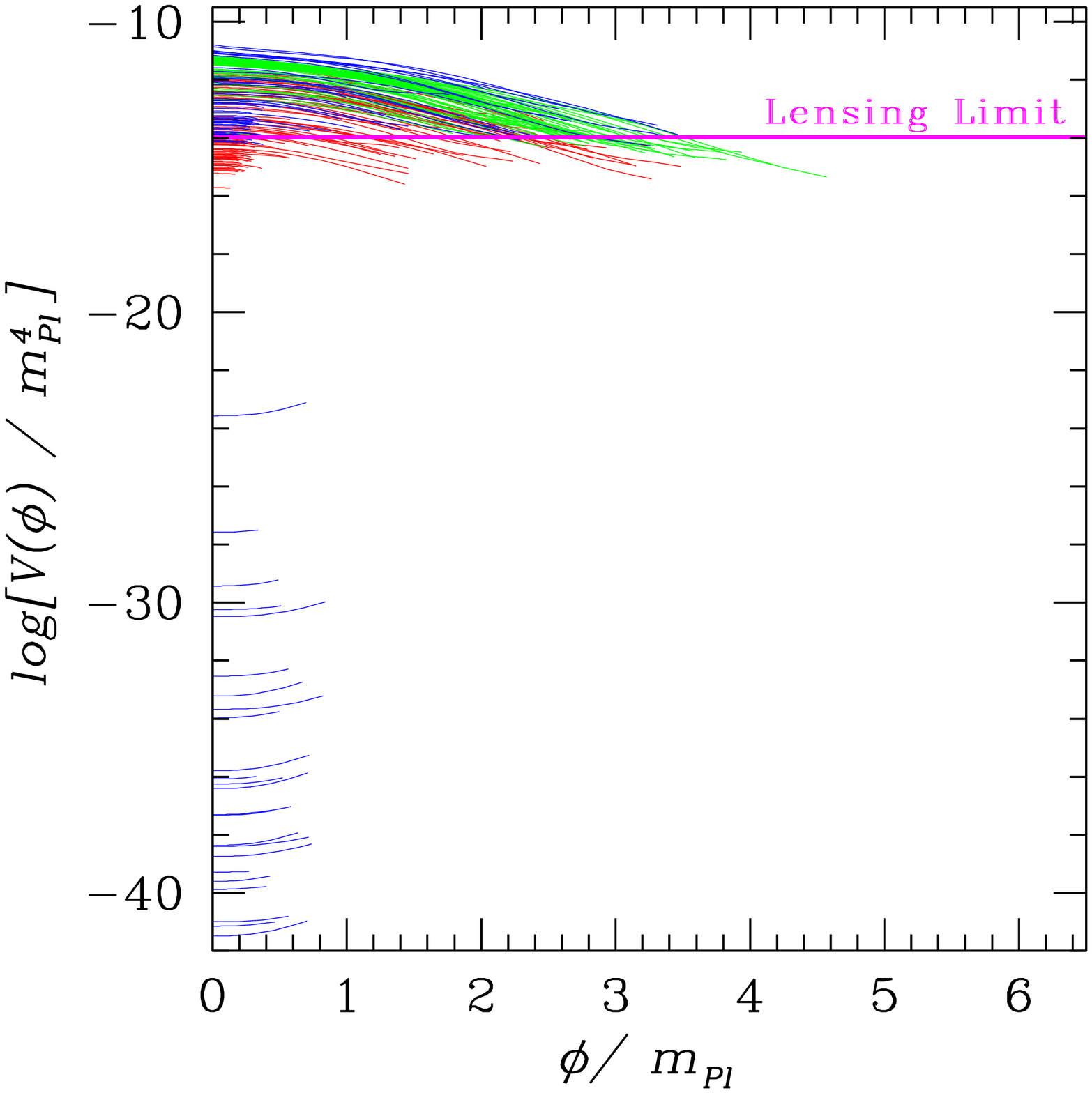}}
\caption{\label{reconWMAP} Potentials consistent with current CMB limits generated
by Monte Carlo reconstruction (color codes are the same as in Figs. 4 and 5).
The horizontal line indicates the lensing limit
of Knox and Song. The energy scale of inflation is not well constrained by
current observations, and could in fact be many orders of magnitude below what
is in principle detectable.}
\end{figure}

What does WMAP tell us about the energy scale of inflation? As one would expect
from the broad range of tensor/scalar ratios consistent with the data, the
height of the potential during inflation is poorly determined. Figure
\ref{reconWMAP} \cite{kkmr} shows an ensemble of potentials consistent with the current
CMB data, generated by Monte Carlo reconstruction \cite{montecarlorecon}.
From Fig. \ref{reconWMAP} we see that a wide range of energy scales are
consistent with existing data, including scales large enough to generate
an (in principle) observable tensor contribution to the CMB anisotropy. We
can also see in concrete form that the width of such potentials is large
enough in Planck units to signal a breakdown of the effective field theory
approximation (\ref{eqeffectivepotential}).

\begin{figure}
\centerline{\includegraphics[width=5.8in]{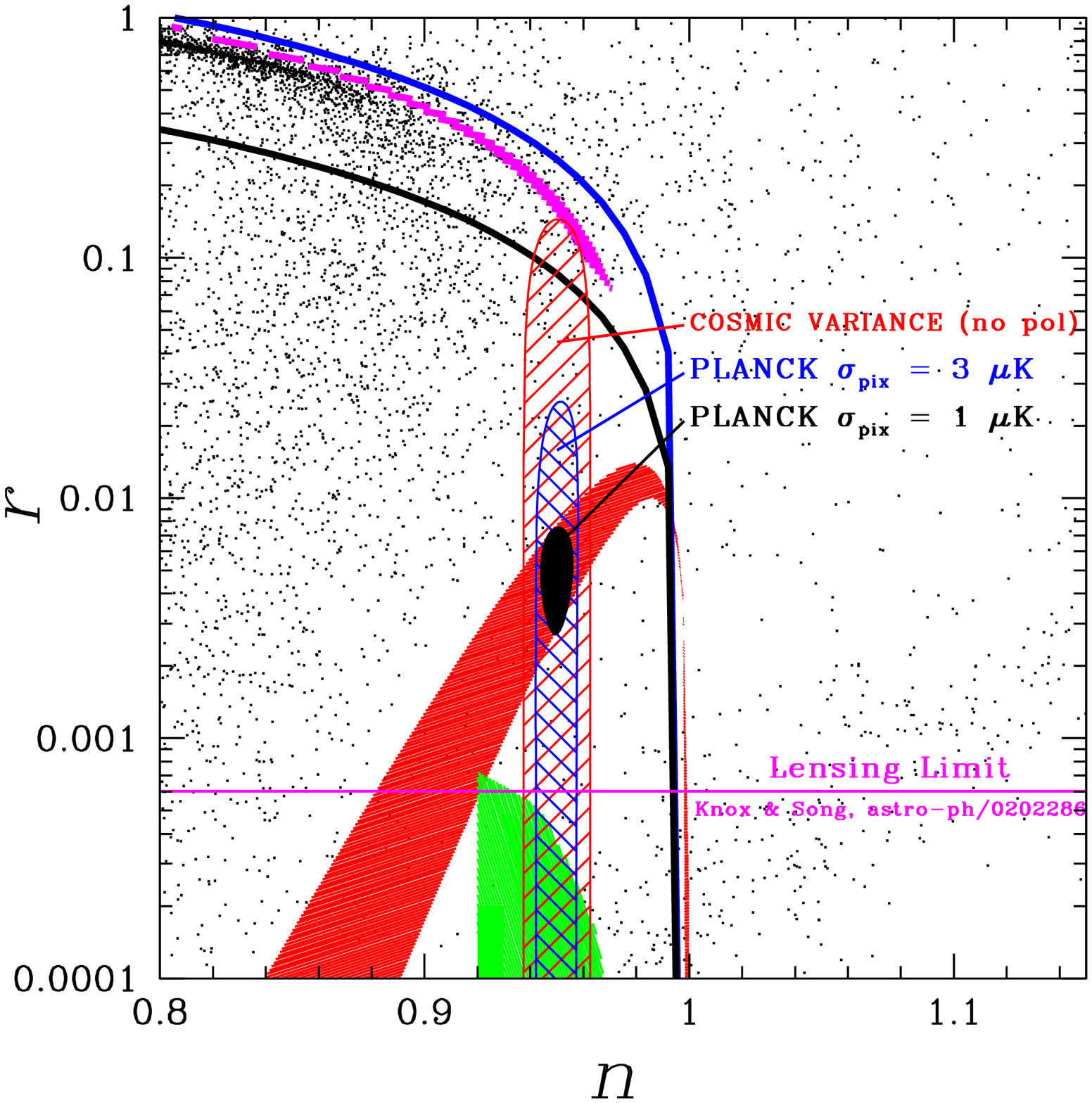}}
\caption{\label{zooplotlog}Inflation models plotted with error bars from
expected and hypothetical measurements. The dots are models generated by
Monte Carlo. The solid areas are the predictions of specific potentials:
$V \propto 1 - (\phi / \mu)^p$ (green, lower area),
$V \propto 1 - (\phi / \mu)^2$ (red, central area),
$V \propto \phi$ (black, third line from top), $V \propto \phi^p$
(cyan, second line from top), and $V \propto \exp(\phi / \mu)$, (blue,
top line). The error bars are for (a) a cosmic-variance limited temperature-only
measurement, (red, outside contour), (b) the Planck Surveyor satellite
including polarization
(blue, middle contour), and (c) a hypothetical measurement with the same
angular resolution as Planck but a factor of three higher sensitivity (black,
inner contour). The horizontal line shows the lensing limit of Knox and
Song.}
\end{figure}

While a tensor/scalar ratio $r > 6 \times 10^{-4}$ is observable in principle,
it is very useful to understand what is going to be possible {\em in practice}
in the near future. Figure \ref{zooplotlog} shows the expected errors from
three experiments, plotted against models on the $n$ vs. $\log{r}$ plane: (1)
a cosmic-variance limited (i.e. ideal) temperature-only CMB map, (2) the
Planck Surveyor satellite, including measurement of CMB polarization \cite{Planck},
and (3) a hypothetical measurement with the same angular resolution as Planck,
but with a factor of three increase in sensitivity \cite{kinney98}. While the
lensing limit on the B-mode is well below the expected sensitivity of these
measurements, a tensor/scalar ratio of $r \sim 0.01$ is well within reach of
presently feasible observations. It may well be possible to do considerably
better.
This level of sensitivity represents an important observational milestone.
A detection of tensor modes in the CMB with $r > 0.01$
will be an indication that the effective field theory expansion
(\ref{eqeffectivepotential}) is not self-consistent, and that therefore inflation
in the early universe is begin driven by some sort of exotic physics,
for example the holographic renormalization group flow proposed by Larsen {\it et al.}
\cite{rginfl}. A non-detection, however, would validate low-energy effective
field theory as a viable tool for inflationary model building, which would
leave the door open for more conventional descriptions of the inflaton, such
as inflation from supersymmetric moduli fields, pseudo-Nambu Goldstone
modes, or other conventional particle physics candidates. (See
Ref. \cite{lythriotto} for a review.)

\section{Conclusions}

Recent CMB observations, especially the results from the WMAP satellite,
represent a sea change in our understanding of the physics of the early
universe. For the first time, it is possible to place meaningful constraints
on theories of the of the universe at its earliest stages, in particular
the inflationary paradigm. If inflation really is the correct model
for the evolution of the early universe and for the generation of primordial
perturbations, a central question is: what was the energy scale of inflation?
Determining the answer to this question requires a measurement of the
primordial gravitational wave (tensor) fluctuations generated during
inflation. The curl, or ``B-mode'' component of CMB polarization is a
particularly sensitive probe of primordial tensor modes, because the
B-mode does not receive contributions from primordial density fluctuations.
However, gravitational lensing by cosmological structure also generates
B-mode polarization, and this foreground cannot be perfectly subtracted.
This places a lower limit on sensitivity to a primordial B-mode and hence
on the amplitude of gravitational wave fluctuations. This can be expressed
in terms of a lower limit on the tensor/scalar ratio $r > 6 \times 10^{-4}$
\cite{knox02}.

Expectations based on the self-consistency of a low-energy effective field
theory description of inflation, however, argue for a very small tensor/scalar
ratio and a corespondingly low energy scale for inflation \cite{lyth96}.
If this theoretical
prejudice is correct, the primordial B-mode will almost certainly be too
small to be detected. Nonetheless, looking at inflationary model building
from the somewhat broader perspective of the flow formalism shows that
there is a substantial population of dynamically viable inflation models with an
observably large tensor/scalar ratio. The expected sensitivity of forthcoming
CMB observations such as the Planck surveyor or (perhaps) dedicated searches
for the B-mode polarization can realistically reach a sensitivity of
$r \sim 0.01$, which is sufficient to distinguish between conventional
effective field theory descriptions of inflation and models based on some
more exotic order parameter. Whether the search for primordial B-mode
polarization in the CMB is ultimately a success or a failure, we will learn
something important about the physics of inflation.

\section*{Acknowledgments}

WHK is supported by ISCAP and  the Columbia University Academic Quality Fund.
ISCAP gratefully acknowledges the generous support of the Ohrstrom Foundation.



\end{document}